\documentclass[12pt]{article}
\usepackage{amsmath, amssymb}

\begin{document}
\title
{Noncommutative Landau problem in graphene: A gauge-invariant analysis with the Seiberg-Witten map}
\author{
{\bf {\normalsize Aslam Halder}$^{a}
$\thanks{aslamhalder.phy@gmail.com}}\\[0.2cm]
%{\bf {\normalsize Sunandan Gangopadhyay}$^{b,c}$\thanks{sunandan.gangopadhyay@gmail.com, sunandan@associates.iucaa.in}}\\[0.2cm]
%{\bf {\normalsize Anirban Saha}$^{a,c}$\thanks{anirban@associates.iucaa.in}}\\[0.2cm]
$^{a}${\normalsize Department of Physics, West Bengal State University, Barasat, Kolkata 700126, India}\\
%$^{b}$ {\normalsize  Department of Theoretical Sciences,}\\
%{\normalsize S. N. Bose National Centre for Basic Science,} \\
%{\normalsize Block-JD, Sector-III, Salt Lake, Kolkata-700106, India}\\[0.2cm]
%$^{c}$ {\normalsize Visiting Associate in Inter University Centre for Astronomy $\&$ Astrophysics (IUCAA),}\\[0.2cm]
%{\normalsize Pune, India}\\[0.3cm]
}

%%%%%%%%%%%%%%%%%%%%%%%%%%%%%%%%%%%%%%%%%%%%%%%%
%\author{
%{\bf {\normalsize Sunandan Gangopadhyay}$^{a,c}
%$\thanks{sunandan.gangopadhyay@gmail.com, sunandan@iiserkol.ac.in, sunandan@associates.iucaa.in}},
%{\bf {\normalsize Aslam Halder }$^{b}$\thanks{aslamhalder.phy@gmail.com}}\\[0.2cm]
%{\bf {\normalsize Anirban Saha}$^{b,c}$\thanks{anirban@associates.iucaa.in}}\\[0.2cm]
%$^{a}$ {\normalsize  Department of Physical Sciences,}\\
%{\normalsize Indian Institute of Science Education and Research Kolkata} \\
%{\normalsize Mohanpur 741246, Nadia, West Bengal, India}\\[0.2cm]
%$^{b}${\normalsize Department of Physics, West Bengal State University, Barasat, Kolkata 700126, India,}\\
%$^{c}$ {\normalsize Visiting Associate in Inter University Centre for Astronomy $\&$ Astrophysics (IUCAA),}\\[0.2cm]
%{\normalsize Pune, India}\\[0.3cm]
%}

\date{}

\maketitle
\begin{abstract}
%Graphene on two dimensional noncommutative (NC) plane in the presence of a constant background magnetic field has been studied. To handel the gauge-invariance issue we start our analysis by a effective massless NC Dirac field theory where we incorporate the Seiberg-Witten (SW) map along with the Moyal star ($\star$) product. The gauge-invariant Hamiltonian of a massless Dirac particle is then computed which is used to study the relativistic Landau problem of graphene on NC plane. Specifically we study the quantum dynamics of a relativistic electron moves on monolayer graphene, in the presence of a constant background magnetic field, in NC plane. We also compute the energy spectrum of the NC Landau system in graphene. The results obtained are corrected by the spatial NC parameter $\theta$.
We investigate the relativistic quantum dynamics of a massless electron in graphene in a two-dimensional noncommutative (NC) plane under a constant background magnetic field. To address the issue of gauge invariance, we employ an effective massless NC Dirac field theory, incorporating the Seiberg-Witten (SW) map alongside the Moyal star ($\star-$) product. Using this framework, we derive a manifestly gauge-invariant Hamiltonian for a massless Dirac particle, which serves as the basis for studying the relativistic Landau problem in NC space. Specifically, we analyze the motion of a relativistic electron in monolayer graphene within this background field and compute the energy spectrum of the NC Landau system. %Our results reveal modifications to the Landau levels due to the spatial NC parameter $\theta$. % highlighting the impact of noncommutativity on the relativistic quantum behavior of graphene. 
The NC-modified energy levels are then used to explore the system’s thermodynamic response. Notably, in the low-temperature limit, spatial noncommutativity leads to a spontaneous magnetization—a distinct signature of NC geometry in relativistic condensed matter systems like graphene.

\end{abstract}

%\begin{abstract}
%We investigate the relativistic quantum dynamics of massless electrons in monolayer graphene within a two-dimensional noncommutative (NC) space under a uniform magnetic field. To ensure gauge invariance, we construct an effective NC Dirac field theory using the Seiberg-Witten map and the Moyal star product, yielding a manifestly gauge-invariant Hamiltonian that incorporates NC corrections perturbatively. Using this Hamiltonian, we solve the relativistic Landau problem and obtain the NC-modified energy spectrum, showing explicit dependence on the spatial NC parameter \( \theta \). Building on this result, we perform a detailed thermodynamic analysis of the system. In particular, we compute the NC-corrected grand potential and magnetization. In the low-temperature limit, spatial noncommutativity leads to a finite spontaneous magnetization even in the absence of an external field—an effect of conceptual significance that may serve as a theoretical signature of NC geometry. Our results highlight the interplay between gauge structure, quantum geometry, and thermodynamic behavior in graphene.
%\end{abstract}

\maketitle

\vskip 1cm
%%%%%%%%%%%%%%%%%%%%%%%%%%%%%%%%%%%%%%%%%%%%%%%%%%%%%%%%%%%%%%%%%%%%%%%%%%%%%%%%%%%%%%%%%%%%%%%%%%%%%%%%%%%%%%%%%%%%%%%%%%%
%%%%%%%%%%%%%%%%%%%%%           SECTION          %%%%%%%%%%%%%%%%%%%%%%%%%%%%%%%%%%%%%%%%%%%%%%%%%%%%%%%%%%%%%%%%%%%%%%%%% %%%%%%%%%%%%%%%%%%%%%%%%%%%%%%%%%%%%%%%%%%%%%%%%%%%%%%%%%%%%%%%%%%%%%%%%%%%%%%%%%%%%%%%%%
\section{Introduction}
Graphene, a two-dimensional arrangement of carbon atoms, has garnered significant research interest across various branches of physics \cite{novo}-\cite{sharapov}. Notably, it is the first truly two-dimensional crystal observed in nature.
Experimental observations indicate that charged fermions near the Dirac points in graphene exhibit relativistic behavior, effectively behaving as massless Dirac quasi-particles \cite{novo}-\cite{peres} . Theoretically, the low-energy electronic excitations in graphene can be described by a (2+1)-dimensional effective massless Dirac field theory \cite{gw}. Of particular interest is the case where such a planar fermionic system interacts with an external gauge field. In this scenario, the dynamics are governed by the massless Dirac field theory, described by the action
\begin{eqnarray}
\label{es}
S=\int{d^4x}\bar{\psi}(x)(\gamma^{\mu}\hat{\pi}_{\mu}){\psi}(x)
\end{eqnarray}
where $\psi$ is the Dirac field, $\pi_{\mu}=p_{\mu}-\frac{e}{c}A_{\mu}(x)$ is the gauge-invariant momentum and $A_{\mu}$ is the U(1) gauge field. 
Interestingly, quantum relativistic effects—typically confined to high-energy physics—manifest at low energies in graphene. This makes graphene an ideal platform for testing quantum field theoretical models, serving as a bridge between condensed matter and high-energy physics.
   %charged fermions which show relativistic behavior \cite{gw}, \cite{shon}. Therefore the quantum dynamics of electron in graphene can thus be described by the Dirac equation which reads 
%\begin{eqnarray}
%\label{evx1}
%i\hbar\frac{\partial\psi}{\partial t}=\hat{H}\psi
%\end{eqnarray}
%where $\hat{H}=v_{F}\vec{\alpha}\cdot\vec{\hat{p}}$ is the Dirac-like Hamiltonian of a massless realivistic electron in graphene, $\psi$ is the 4-component spinor describes the electronic state around the Dirac points $k$ and $k'$, $v_{F}=10^6$ is the Fermi velocity which plays the role of the speed of light in vacuum, $\hat{p}$ is the momentum operator of the electron and $\alpha$ is the well known Dirac matrix.
%Interestingly, quantum relativistic phenomena which is typically a high energy physics occurs at low energy in graphene however.  
%This relativistic behavior of graphene makes it an ideal candidate for the test of quantum field theoretical models. Therefore graphene is a bridge between condensed matter and high energy physics.

%On the other hand there has been an upheaval in investigating the physics of systems living in a NC space-time in the last two decades 
%Conversely, the past two decades have seen a surge in research focused on the physics of systems existing within a noncommutative space-time \cite{sw}-\cite{szabo}.
Conversely, over the past two decades, there has been a surge in research exploring the physics of systems within a noncommutative space-time \cite{sw}-\cite{szabo}.
NC space-time arises in string theory with $D$-branes in the background of Neveu-Schwarz fields \cite{acny}-\cite{braga}. It is found that the $D$-brane world volume becomes an NC space and one can arrive at a low energy effective field theory in the point particle limit where the string length goes to zero. This gives rise to a noncommutative quantum field theory (NCQFT) \cite{sw},\cite{scho},\cite{szabo},\cite{carroll}-\cite{bal} where the NC coordinate algebra
\begin{eqnarray}
\label{eal}
\left[\hat{X}^{\mu},\hat{X}^{\nu}\right]=i\theta^{\mu\nu}~,
\end{eqnarray}
with the constant anti-symmetric tensor $i\theta^{\mu\nu}$, leads to an uncertainty in the space-time geometry and the notion of a space-time point is replaced by a Planck cell. Besides these, various theories of quantum gravity has also led to NC geometry \cite{suss}-\cite{mof}. Remarkably, such NC geometry can also manifest in low-energy condensed matter systems. A prominent example is the Landau problem, where the electron is no longer a point-like particle but is instead localized at the scale of the magnetic length, leading to significant conceptual changes. This has sparked considerable interest in studying the Landau system in NC space \cite{jellal}-\cite{ahsg2}.

%an electron moving in a plane under a strong perpendicular magnetic field exhibits an effective noncommutative structure: its guiding center coordinates no longer commute, resembling the NC algebra of space. This profound analogy has led to considerable interest in revisiting the Landau system within NC frameworks, thereby offering a unique interplay between quantum mechanics, field theory, and geometry.

%Surprisingly, in direct analogy with the string theoretical case, noncommuting coordinates can arise in a simple quantum mechanical system. To be specific, the dynamics of charged particles in the presence of electromagnetic fields, popularly known as the Landu problem \cite{landau}
 %of charged particle in background EM field, popularly known as the Landau problem \cite{landau}, 
%when the lowest Landau Level is partially filled. Interestingly the conceptual foundation changes drastically in the NC scenario, e.g, the electron is no longer a point like particle and can at best be localized at the scale of the magnetic length. 
 %Therefore the study of Landau system in NC space has drawn numerous interest in the literature \cite{jellal}-\cite{ahsg2}.  

%Interestingly, NC coordinates can emerge in simple quantum systems, analogous to string theory. A notable example is the Landau problem \cite{landau} where the electron is no longer a point-like particle but is instead localized at the scale of the magnetic length, leading to significant conceptual changes. This has sparked considerable interest in studying the Landau system in NC space \cite{jellal}-\cite{ahsg2}.

The low-energy limit of NCQFT leads to noncommutative quantum mechanics (NCQM), a framework that has been extensively explored in the literature \cite{jellal}-\cite{gov}. In this setting, the fundamental algebra (which is not the most general form) governing the canonical pairs $(\hat{X}_{i},\hat{P}_{i})$ in a two-dimensional NC plane is given by
\begin{eqnarray}
\label{e420}
\left[\hat{X}_{i},\hat{X}_{j}\right]=i\theta_{ij} = i\theta \epsilon_{ij} ~ ; \quad \left[\hat{X}_{i},\hat{P}_{j}\right] = i\hbar\delta_{ij} ~ ;\quad \left[\hat{P}_{i},\hat{P}_{j}\right]=0~.
\end{eqnarray}
Here $\theta$ is the spatial NC parameter and is antisymmetric in the indices $i,j$ as $\theta^{ij}=\theta \epsilon^{ij}$, where $\epsilon^{ij}=-\epsilon^{ji}, (\epsilon^{12}=1)$.
The usual approach in the literature to deal with such problems is to form an equivalent commutative description of the NC theory by employing some transformation which relates the NC operators $\hat{X}_{i}$, $\hat{P}_{i}$ to ordinary commutative operators $\hat{x}_{i}$, $\hat{p}_{i}$ satisfying the usual Heisenberg algebra
\begin{eqnarray}
\left[\hat{x}_{i}, \hat{p}_{j}\right]=i\hbar \delta_{ij}~; \quad \left[\hat{x}_{i} \, , \, \hat{x}_{j}\right]=0= \left[\hat{p}_{i},\hat{p}_{j}\right].
\label{cAlgebra}
\end{eqnarray} 
%%%%%%%%%%%%%%%%%%%%%%%%%%%%%%%%%%%%%%%%%%%%%%%%%%%%%%%%%%%%%%%%%
 %All the studies of NCQM give rise to NC corrections to the standard results. Obviously, due to the extreme smallness of the NC parameter $\theta$ it is for sure that its effects will not be observed in the near future. However, there is a nice motivation of studying quantum systems in NC space-time. The physical system in the usual framework can be linked to a noninteracting theory in the NC framework thereby presaging at a possible duality between the two systems

Studies on NCQM consistently reveal NC corrections to standard results. Although the smallness of the NC parameter $\theta$ renders its effects unobservable in the near future, exploring quantum systems in NC space-time remains valuable. Notably, mapping a physical system to a noninteracting theory in the NC framework suggests a possible duality \cite{dayijellal}-\cite{gov} and provides bounds on $\theta$ \cite{carroll}, \cite{galileo}, \cite{AHepl}, further motivating the study of NC space-time.

Recent studies on relativistic NCQM \cite{AHepl}-\cite{PLB} have sparked interest in (2+1)-dimensional noncommutativity in graphene, a natural toy model for exploring potential traces of (3+1)-dimensional noncommutativity. However, many works in this area of the literature \cite{bertolami}, \cite{catrina}, \cite{vsantos}, \cite{epj} fail to address NC gauge invariance, leading to a non-gauge-invariant formulation of the Hamiltonian, which consequently affects physical results such as the energy spectrum and susceptibility.
This raises the question of whether a gauge-invariant formulation is possible. In our recent work \cite{AHepl}, we resolve the issue of non-gauge-invariance for a Dirac particle coupled with a gauge field by adopting a manifestly gauge-invariant approach using the SW map \cite{sw} along with the star ($\star-$) product \cite{mezin}, ensuring a consistent description. In the present work we follow the same prescription to study the Landau problem of graphene in NC plane.

In this work, we explore the dynamics of a massless NC spinor field interacting with a U$(1)_{\star}$ background gauge field in NC space. By employing the SW map and expanding the star  ($\star-$) product to first order in the NC parameter $\theta$, we derive a manifestly U(1) gauge-invariant commutative action, where NC effects emerge as perturbative corrections. From this framework, we obtain the $\theta$-modified Dirac equation and identify the corresponding Dirac Hamiltonian for a relativistic electron in graphene, ensuring gauge invariance is preserved. Utilizing this Hamiltonian, we investigate the relativistic Landau problem in NC space, revealing how the position and mechanical momentum operators evolve over time, incorporating NC corrections at the equation-of-motion level. We then move on to compute the energy spectrum, uncovering the imprint of spatial noncommutativity through modifications governed by $\theta$.  Building upon the modified Landau spectrum, we further analyze the thermodynamic properties of the system—specifically the grand potential, magnetization—demonstrating how noncommutativity introduces corrections to these quantities, particularly in the low-temperature regime.  Notably, we uncover a striking consequence of noncommutativity: the emergence of a spontaneous magnetization, which serves as a distinct signature of NC geometry in relativistic condensed matter systems.

The present article is organized as follows. In Section 2, we provide the theoretical framework, introducing the dynamics of a massless relativistic charged fermion on a NC plane under a constant background electromagnetic (EM) field. The implementation of the SW map to maintain NC gauge invariance is also discussed. In Section 3, we investigate the relativistic Landau problem for graphene in NC space, focusing on the time evolution of the position and mechanical momentum operators, and derive the modified energy spectrum. In Section 4, we analyze the thermodynamic properties of the NC Landau system in graphene. Finally, in Section 5, we summarize our key results and present concluding remarks.

%The present article is organized as follows. In section 2, we give the basic outline which involves introducing the problem of a massless relativistic charged fermion moving on NC plane in the presence of a constant background EM field. The SW map in the context of NC gauge-invariance is incorporated here. In section 3, relativistic Landau problem of graphene on NC plane is studied. Specifically, we compute the time evolution of the position and mechanical momentum operators. We also compute the energy spectrum of the relativistic NC Landau system in graphene. In section 4, we study the thermodynamics of the Landau system in graphene. Finally we summarize our findings in section 5.

%%%%%%%%%%%%%%%%%%%%%%%%%%%%%%%%%%%%%%%%%%%%%%%%%%%%%%%%%%%%%%%%%
%%%%%%%%%%%%%%%%%%%%%%%%SECTION%%%%%%%%%%%%%%%%%%%%%%%%%%%%%%%%%%%%%%
\section{Gauge-invariant formulation in noncommutative space}%Gauge-invariance algorithm in noncommutative space} %of the problem of relativistic charged particle coupled to background electromagnetic field in noncommutative space}
%In this section, we want to give a brief overview about the issue of gauge-invariance for the problem of a massless charged spin-$1/2$ particle coupled with EM field in NC space. At first we demonstrate the problem by starting with NCQM description directly where we simply replace the ordinary
%product rule among the operators in the quantum mechanical Hamiltonian by Moyal Star ($\star$) product. This yields a matter of non-gauge-invariance. Then we handel this non-gauge-invariance issue by a NC field theoretic approach where we use the SW map along with the Star ($\star$) product and consequently we reach a manifestly gauge-invariant commutative equivalent description of the NC problem.

This section explores the fundamental challenge of preserving gauge invariance in the dynamics of a massless charged spin-$1/2$ particle interacting with an EM field in NC space. We first illustrate the issue by adopting the framework of NCQM, wherein the conventional operator multiplication in the quantum Hamiltonian is supplanted by the star ($\star-$) product. However, this naive extension disrupts gauge symmetry, rendering the formulation inherently non-invariant. To remedy this, we transition to a field-theoretic perspective, leveraging the SW map in conjunction with the star ($\star-$) product. This approach systematically restores gauge invariance by embedding NC effects within a perturbative expansion, ultimately yielding a manifestly gauge-invariant commutative equivalent of the NC system.
\subsection{Challenges of gauge-non-invariance}
%In this section we want to study the relativistic Landau levels of graphene in NC space. To do so we consider a monolayer graphene sheet in two dimensional NC plane in the presence of a background EM field. 
%We consider a relativistic massless charged spin-$1/2$ particle of charge $e$ moving on NC plane in the presence of a background EM field. 
%The quantum dynamics of this particle can be described by the generalization of the standard Dirac equation for such particle in
%the commutative space to the NC space as 

We consider a relativistic, massless, charged spin-$1/2$ particle with charge $e$ moving on a NC plane under the influence of a background EM field. 
The quantum dynamics of this system are described by extending the conventional Dirac equation to its noncommutative (NC) counterpart, where operator multiplication is replaced by the Moyal star ($\star-$) product. This product, defined for two functions $f(X)$, $g(X)$, defined as
\begin{eqnarray}
\label{ev03m}
f(X)\star g(X)=f(x)exp\left\{\frac{i}{2}\theta^{\mu\nu}\overleftarrow{\partial_{\mu}}\overrightarrow{\partial_{\nu}}\right\}g(x),
\end{eqnarray} 
encodes the noncommutative structure of space-time. Incorporating this deformation, the NC Dirac equation takes the form
\begin{eqnarray}
\label{evxm}
i\hbar\frac{\partial\Psi(X)}{\partial t}=\hat{H}(X, P)\star\Psi(X)
\end{eqnarray}
where the Hamiltonian is given by
\begin{eqnarray}
\label{evxy}
\hat{H}(X,P)=c\vec{\alpha}\cdot\left(\vec{\hat{P}}-\frac{e}{c}\vec{\hat{\mathcal{A}}}(X)\right)+e\phi(X)~.
\end{eqnarray}
Here, $\Psi(X)$ denotes the $4$-component NC Dirac spinor, while $\hat{P}$, ${\mathcal{A}}(X)$ and $\phi(X)$ are the generalized momentum, EM vector potential and scalar potential respectively in NC space.

%Here $\theta$ is the spatial NC parameter and is antisymmetric in the indices $i,j$ as $\theta^{ij}=\theta \epsilon^{ij}$, where $\epsilon^{ij}=-\epsilon^{ji}, (\epsilon^{12}=1)$.

Applying (\ref{ev03m}) to first order in spatial NC parameters $\theta^{ij}$ in eq. (\ref{evxm}) we get  the NC corrected (up to first order in $\theta$) Dirac equation as
\begin{eqnarray}
\label{ej467}
i\hbar\frac{\partial\Psi(x)}{\partial t}&=&\hat{H}(x, p)\Psi(x)+\frac{i}{2}\theta^{jk}\partial_{j}\hat{H}(x, p)\partial_{k}\Psi(x)\nonumber\\
&=&\hat{H}(x, p)\Psi(x)+\frac{i}{2}\theta^{jk}\partial_{j}\left\{c\alpha_{j}\left(\hat{p}_{j}-\frac{e}{c}\hat{\mathcal{A}}_{j}(x)\right)+e\phi(x)\right\}\partial_{k}\Psi(x)~.
\end{eqnarray}
From the above equation we can easily identify the $\theta$ modified Dirac Hamiltonian of the system as\footnote{$\theta^{i}=\frac{1}{2}\epsilon_{ijk}\theta^{jk}$}
\begin{eqnarray}
\label{es04}
\hat{H}=c\vec{\alpha}\cdot\left(\vec{\hat{p}}-\frac{e}{c}\vec{\hat{\mathcal{A}}}(x)\right)+e\phi(x)+\frac{e}{2\hbar}\left\{\vec{\bigtriangledown}\left(c\vec{\alpha}\cdot\vec{\hat{\mathcal{A}}}(x)-\phi(x)\right)\times\vec{\hat{p}}\right\}\cdot\vec{\theta}~.
\end{eqnarray}

%The above Hamiltonian is not manifestly gauge invariant due to the presence of gauge dependent term in the $\theta$ correction portion. Therefore this Hamiltonian is not suitable for studying the relativistic Landau problem of graphene. 
%In \cite{bertolami}, \cite{catrina} this gauge symmetry breaking issue is appeared. To avoid this difficulty the authors renounce the spatial noncommutativity and consider only momentum noncommutativity in their work. This matter shed light on the fact that when NC system is coupled with a NC gauge field, the quantum mechanical treatment (i.e., considering only star product) of the problem is an incomplete description in conformity with NC gauge invariance.

The Hamiltonian derived above lacks manifest gauge invariance due to the presence of explicitly gauge-dependent terms in the $\theta$ corrected sector. This inherent gauge symmetry breaking renders the Hamiltonian unsuitable for a consistent formulation of the relativistic Landau problem in graphene. Previous works, such as \cite{bertolami}, \cite{catrina}, \cite{epj} have encountered this issue and attempted to circumvent it by entirely discarding spatial noncommutativity while retaining only momentum-space noncommutativity in their analyses.

This observation highlights a crucial limitation: when a NC system is coupled to an NC gauge field, a purely quantum mechanical treatment—wherein the star ($\star-$) product is directly applied to the Hamiltonian—proves to be an incomplete framework from the perspective of NC gauge invariance. The breakdown of gauge symmetry in this approach indicates that a more refined theoretical formulation is necessary to incorporate NC effects consistently while preserving the fundamental gauge structure of the theory.

\subsection{Gauge-invariant noncommutative massless Dirac field theory} 
%To tackle the issue of gauge-non-invariance we want to use a field theoretic approach, has been used in \cite{Adorno}, in the problem where we will generalize the system from commutative space to NC space and finally sticking to a commutative description of the NC system. The NC generalization is done through replacing the ordinary commutative spinor field $\psi$ and U(1) gauge field $A_{\mu}$ in the action (\ref{es}) by the  NC spinor field ${\Psi}(X)$ and U$(1)_{\star}$ gauge field ${\mathcal{A}}^{\mu}(X)=({\mathcal{A}}^{0}, {\mathcal{A}}^{i});~i=1,2,3$. Also the ordinary product should be replaced by the Moyal star product. The corresponding U$(1)_{\star}$ gauge invariant action is 

To systematically restore gauge invariance in the NC framework, we adopt a field-theoretic approach, as explored in \cite{AHepl}. Instead of applying the Moyal star product directly to the quantum Hamiltonian, which disrupts gauge symmetry, we reformulate the problem by embedding the system within a gauge-invariant NC field theory. This is achieved by first generalizing the commutative Dirac field theory to its NC counterpart and then utilizing the SW map to construct a commutative equivalent description that retains the effects of noncommutativity in a perturbative manner. 

The NC generalization proceeds by replacing the standard commutative spinor field $\psi$ and U(1) gauge field $A_{\mu}$ in the action (\ref{es}) with their NC counterparts: the NC spinor field ${\Psi}(X)$ and U$(1)_{\star}$ gauge field ${\mathcal{A}}^{\mu}(X)=({\mathcal{A}}^{0}, {\mathcal{A}}^{i});~i=1,2,3$.
Furthermore, ordinary operator multiplication is replaced by the star ($\star-$) product, ensuring a consistent deformation of the gauge structure. Consequently, the gauge-invariant action for a massless Dirac field in NC space takes the form
\begin{eqnarray}
\label{eurm}
S&=&\int{d^4x}\bar{\Psi}(X)\star(\gamma^{\mu}\hat{\Pi}_{\mu})\star{\Psi}(X)
%\hat{\mathcal{L}}&=&{\Psi}^{\dagger}\star(\gamma^{\mu}\hat{\Pi}_{\mu})\star{\Psi},\nonumber\\
%\hat{\Pi}_{\mu}&=&\hat{P}_{\mu}-\frac{e}{c}\hat{\mathcal{A}}_{\mu}(X)\nonumber~.%,~\hat{p}_{\mu}=-i\hbar \partial_{\mu},\nonumber\\
%{\mathcal{A}}^{\mu}&=&({\mathcal{A}}^{0}, {\mathcal{A}}^{i});~i=1,2,3~.\nonumber
\end{eqnarray}
where $\Pi_{\mu}=P_{\mu}-\frac{e}{c}\mathcal{A}_{\mu}(X)$ is the NC generalization of the commutative mechanical momentum $\pi_{\mu}=p_{\mu}-\frac{e}{c}{A}_{\mu}(x)$. 
To establish a physically meaningful and gauge-consistent formulation, we express the NC gauge field ${\mathcal{A}}^{\mu}(X)$ and the NC spinor field ${\Psi}(X)$
in terms of their commutative counterparts using the SW map, truncated to first order in the NC parameter $\theta$ \cite{sw}
%The NC gauge field ${\mathcal{A}}_{\mu}$ and the NC spinor field ${\Psi}$ are expressed in terms of ordinary commutative gauge field $A_{\mu}=(A_{o}, A_{i})$ and commutative spinor field $\psi$ by the SW map (up to first order in $\theta$) \cite{sw}
\begin{eqnarray}
\label{e159}
{\mathcal{A}}_{\mu}&=&{A}_{\mu}+\frac{e}{2\hbar c}\theta^{\alpha\beta}{A}_{\alpha}(\partial_{\beta}{A}_{\mu}+F_{\beta\mu})\\
{\Psi}&=&\psi+\frac{e}{2\hbar c}\theta^{\alpha\beta}{A}_{\alpha}\partial_{\beta}\psi~.\nonumber
\end{eqnarray}
%Applying the above map and subsequently using the Moyal star product in the action (\ref{eur}), we compute the $\theta$ modified commutative equivalent action as
These transformations enable us a systematic derivation of the commutative equivalent action corresponding to (\ref{eurm}) which reads 
\begin{eqnarray}
\label{e163vx}
S^{\theta}&=&\int d^4x\bar{\psi}(x)\left\{\gamma^{\mu}\left[\left(1+\frac{e}{4c\hbar}\theta^{\alpha\beta}F_{\alpha\beta}\right)\hat{\pi}_{\mu}- \frac{e}{2c\hbar}\theta^{\alpha\beta}F_{\alpha\mu}\hat{\pi}_{\beta}\right]\right\}\psi(x)~.
\end{eqnarray}
%The above action is manifestly U(1) gauge invariant due to the appearance of the of commutative gauge
%field strength tensor $F_{\alpha\beta}$ and commutative momentum field $\pi_{\mu}$. It should be noted that incorporating of the SW map along with the star product in our work confirm the gauge-invariance of the above action.  
The resulting action maintains manifest U(1) gauge invariance, as it explicitly depends on the commutative gauge field strength tensor $F_{\alpha\beta}$
and the gauge-covariant momentum $\pi_{\mu}$. Crucially, the implementation of the SW map, along with the star ($\star-$) product, systematically incorporates NC corrections while ensuring gauge symmetry within the first-order approximation in $\theta$.

To obtain the equation of motion from the gauge-invariant action (12), we apply the Euler-Lagrange equation with respect to $\bar{\psi}$, which yields
\begin{equation}
\gamma^\mu \left[ \left(1 + \frac{e}{4\hbar c} \theta^{\alpha\beta} F_{\alpha\beta} \right) \hat{\pi}_\mu - \frac{e}{2\hbar c} \theta^{\alpha\beta} F_{\alpha\mu} \hat{\pi}_\beta \right] \psi = 0. \tag{13}
\end{equation}
Separating the temporal and spatial components and rearranging terms, we express this in Schrödinger form as\footnote{The full derivation is provided in Appendix A.}
\begin{equation}
i\hbar \frac{\partial \psi}{\partial t} = \hat{H}_{\text{NC}} \psi, \tag{18}
\end{equation}
with the noncommutative Hamiltonian given by
\begin{equation}
\hat{H}_{\text{NC}} = \hat{H} + \Delta \hat{H}_\theta, \tag{19}
\end{equation}
where
\begin{equation*}
\hat{H} = c\, \vec{\alpha} \cdot \left( \hat{\vec{p}} - \frac{e}{c} \vec{\hat{A}} \right) + eA_0,
\quad
\Delta \hat{H}_\theta = \frac{e}{2\hbar} \left[ (\vec{E} \times \hat{\vec{\pi}}) \cdot \vec{\theta} + (\vec{\theta} \times (\vec{\alpha} \times \vec{B})) \cdot \hat{\vec{\pi}} \right].
\end{equation*}
%This Hamiltonian, obtained from the gauge-invariant action, consistently incorporates the leading-order effects of spatial noncommutativity while preserving $U(1)$ gauge symmetry. 
Since the Hamiltonian $\hat{H}_{\rm NC}$ originates from the manifestly gauge-invariant action (\ref{e163vx}), its gauge invariance is inherently preserved at all levels of the theory. Notably, the absence of explicit gauge-dependent terms in $\hat{H}_{\rm NC}$ stands in stark contrast to the non-gauge-invariant Hamiltonian (9), which suffers from a breakdown of gauge symmetry. This ensures the consistency of our framework, allowing us to employ this gauge-invariant Hamiltonian in the subsequent analysis of the relativistic Landau problem in graphene. The presence of NC corrections encapsulated within $\theta$ introduces subtle modifications to the relativistic quantum dynamics, which will be explored in the following sections.

%%%%%%%%%%%%%%%%%%%%%%%%%%%%%%%%%%%%%%%%%%%%%%%%%%%%%%%%%%%%%%%%%%%
%%%%%%%%%%%%%%%%%%%%%%%%%%%%Section%%%%%%%%%%%%%%%%%%%%%%%%%%%%%%%%%%%%%%%
%%%%%%%%%%%%%%%%%%%%%%%%%%%%%%%%%%%%%%%%%%%%%%%%%%%%%%%%%%%%%%%%%%%

\section{Relativistic Landau quantization of graphene in noncommutative space}
We are now in a position to investigate the relativistic Landau problem of graphene in a NC plane, specifically the quantum dynamics of a massless relativistic electron confined to a monolayer graphene sheet within a two-dimensional NC plane, subjected to a homogeneous background magnetic field oriented perpendicular to the plane. To facilitate this analysis, we consider the magnetic field directed along the $z$-axis, ensuring that the system remains constrained within the NC \( x \)-\( y \) plane.  
With this setting and by a little manipulation of the gauge-invariant Hamiltonian (14), we express the Hamiltonian governing the relativistic Landau system in graphene as\footnote{As the electron is confined to a plane, spatial noncommutativity is introduced as $\theta_{ij} = \epsilon_{ij} \theta$,  $\left( i,j = 1,2\right)$. In studying the Landau problem, we consider only the magnetic field, assuming the electric field to be absent.}  
\begin{eqnarray}
\label{e4x9t}
\hat{H}_{\rm NC} &=& v_F \vec{\alpha} \cdot \left( \hat{\vec{p}} - \frac{e}{c} \hat{\vec{A}} \right) \left( 1 + \frac{eB\theta}{2\hbar c}\right)~.%\nonumber\\
%&=& v_F \boldsymbol{\alpha} \cdot \hat{\boldsymbol{\pi}} \left( 1 + \frac{eB\theta}{2\hbar c} \right)
\end{eqnarray}
Expanding in component form, this becomes  
\begin{eqnarray}
\label{e163kj}
   \hat{H}_{\rm NC} = v_F \left[ \alpha_x \left( \hat{p}_x + \frac{eB \hat{y}}{2c} \right) + \alpha_y \left( \hat{p}_y - \frac{eB \hat{x}}{2c} \right) \right] \left( 1 + \frac{eB\theta}{2\hbar c} \right),
\end{eqnarray}
where we have fixed the symmetric gauge choice for the vector potential $\vec{A}\equiv\left(-\frac{By}{2},\frac{Bx}{2},0\right)$  
and replace the speed of light in vacuum \( c \) with the Fermi velocity \( v_F \approx 10^6 \) m/s, which characterizes the relativistic massless quasiparticles propagating within the honeycomb lattice of graphene. It should be noted that the manifest gauge invariance of the commutative equivalent action (\ref{e163vx}) confirms that our theory remains gauge-invariant at all subsequent levels (i.e., the equation of motion, the energy spectrum, etc.). Therefore, we can choose this gauge without loss of generality.

%%%%%%%%%%%%%%%%%%%%%%%%%%%%%%%%%%%%%%%%%%%%%%%%%%%%%%%%%%%%%%%%%
\subsection{Equation of motion of the relativistic electron in graphene}
Having established the gauge-invariant framework, we now proceed to examine the relativistic dynamics of a massless electron in graphene under a background magnetic field in NC space.
The time evaluation of the position operator (we deliberately refer this quantity as the velocity operator in the reminder) is computed by using the Hamiltonian (\ref{e163kj}) as 
\begin{eqnarray}
\label{e4x} 
\vec{v}&=&\frac{i}{\hbar}[\hat{H}_{\rm NC}~, ~\hat{i}\hat{x}+\hat{j}\hat{y}]\nonumber\\
%&=&\frac{i}{\hbar}\left[v_{F}\left\{\alpha_{x}\left(\hat{p}_{x}+\frac{eB\hat{y}}{2c}\right)+\alpha_{y}\left(\hat{p}_{y}-\frac{eB\hat{x}}{2c}\right)\right\}\left(1+\frac{eB\theta}{2\hbar c}\right)%\right.\nonumber\\
%&&\left.-eE\hat{x}+\frac{eE\theta}{2\hbar}\left(\hat{p}_{y}-\frac{eB\hat{x}}{2c}\right)+\beta mc^2~,
%,~\hat{i}\hat{x}+\hat{j}\hat{y}\right]\nonumber\\
&=&v_{F}\vec{\alpha}\left(1+\frac{eB\theta}{2\hbar c}\right)~.
\end{eqnarray}
The time-evolution of the gauge-invariant
mechanical momentum operator $\vec{\hat{\pi}}$ (which is analogue to the classical Lorentz force) in NC space is obtained as
\begin{eqnarray}
\label{e4xgy} 
\frac{d}{dt}(\vec{\hat{\pi}})\equiv\vec{F}^{\rm NC}&=&\frac{i}{\hbar}[\hat{H}_{\rm NC}~,~\vec{\hat{\pi}}]\nonumber\\
&=&\frac{i}{\hbar}[\hat{H}_{\rm NC}~,~\hat{i}\hat{\pi}_{x}+\hat{j}\hat{\pi}_{y}]\nonumber\\
%&=&\frac{i}{\hbar}\left[v_{F}\left\{\alpha_{x}\left(\hat{p}_{x}+\frac{eB\hat{y}}{2c}\right)+\alpha_{y}\left(\hat{p}_{y}-\frac{eB\hat{x}}{2c}\right)\right\}\left(1+\frac{eB\theta}{2\hbar c}\right)\right.\nonumber\\
%&&\left.~,~\hat{i}\left(\hat{p}_{x}+\frac{eB\hat{y}}{2c}\right)+\hat{j}\left(\hat{p}_{y}-\frac{eB\hat{x}}{2c}\right)\right]\nonumber\\
&=&\frac{1}{c}e\vec{v}\times\vec{B}^{\rm NC}~.
\end{eqnarray}
From the above expression, we observe that the relativistic electron in graphene experiences the usual Lorentz force law. However, it perceives an effective magnetic field that includes a NC correction, given by
\begin{eqnarray}
\label{e4xgy} 
{B}^{\rm NC}=B\left(1+\frac{eB\theta}{2\hbar c}\right)~.
\end{eqnarray} 
% Notice that the NC effect arise completely at the equation of motion level. Since equation of motion determines the experimentally observable dynamics of any
%physical system so we can safely assume that any observable NC quantum mechanical effect displayed by our
%system can be captured by replacing the standard velocity and Lorentz force by the effective velocity and Lorentz force. This is a crucial
%result of our paper. At $\theta=0$ limit, the above expressions for the velocity and the
%Lorentz force are exactly match with their commutative forms.
It is important to note that NC effects manifest exclusively at the equation-of-motion level. Since the equation of motion governs the experimentally observable dynamics of any physical system, we infer that any measurable NC quantum mechanical effects can be effectively captured by substituting the standard velocity and Lorentz force with their NC-corrected counterparts. This constitutes a key result of our study. In the limit $\theta=0$, the modified expressions for velocity and the Lorentz force precisely reduce to their commutative counterparts, ensuring consistency with conventional results.

%%%%%%%%%%%%%%%%%%%%%ENERGY SPECTRUM%%%%%%%%%%%%%%%%%%%%%%%%%%%%%%%%%%%%%
\subsection{Energy spectrum of the noncommutative Landau System in Graphene}
To derive the energy spectrum of the relativistic Landau system in graphene within the NC plane, we recast the gauge-invariant Hamiltonian (\ref{e4x9t}) as  

\begin{equation}
\hat{H} = v_F \vec{\alpha} \cdot \hat{\vec{\pi}}  
\left( 1 + \frac{e B \theta}{2 \hbar c} \right).
\end{equation}

Adopting the standard Dirac representation, the Hamiltonian assumes a block-diagonal form,

\begin{equation}
\hat{H} =
v_F
\begin{pmatrix}
\vec{\sigma} \cdot \tilde{\hat{\vec{\pi}}} & 0 \\
0 & -\vec{\sigma} \cdot \tilde{\hat{\vec{\pi}}}
\end{pmatrix}
=
v_F
\begin{pmatrix}
\hat{H}_K & 0 \\
0 & \hat{H}_{K'}
\end{pmatrix}.
\end{equation}

Here, $\tilde{\pi}=\hat{\pi} \left( 1 + \frac{e B \theta}{2 \hbar c} \right)$ is the NC-modified mechanical momentum operator %is defined as  
%\begin{equation}
%\tilde{\hat{\vec{\pi}}} = \hat{\vec{\pi}}  
%\left( 1 + \frac{e B \theta}{2 \hbar c} \right),
%\end{equation}
which introduces a perturbative correction to the conventional relativistic Landau problem. The Hamiltonian naturally decomposes into two independent sectors associated with the Dirac points \( K \) and \( K' \) in graphene’s Brillouin zone. For the \( K \)-valley, the effective Hamiltonian simplifies to  

\begin{equation}
\hat{H}_K = v_F \vec{\sigma} \cdot \tilde{\hat{\vec{\pi}}}.
\end{equation}

%This formulation preserves gauge invariance while systematically incorporating noncommutative effects. The NC correction manifests as a momentum rescaling, effectively modifying the relativistic Landau level structure. The ensuing spectral analysis will quantify these modifications and provide insights into the role of spatial noncommutativity in graphene’s quantum dynamics.
%%%%%%%%%%%%%%%%%%%%%%%%%%%%%%%%%%%%%%%%%%%%%%%%%%%%%%%%%%%%%%%%%%%%%

To determine the energy eigenvalues of the above Hamiltonian, we begin by computing the commutator between the two components of the NC-modified mechanical momentum operator 
 $\tilde{\vec{\pi}}$ operator, which yields 
\begin{eqnarray}
\label{eo0w32}
[\tilde{\pi}_{x}~,~\tilde{\pi}_{y}]%&=&\left(1+\frac{eB\theta}{2\hbar c}\right)^2\left[\hat{p}_{x}-\frac{e}{c}\hat{A}_{x}~,~\hat{p}_{y}-\frac{e}{c}\hat{A}_{y}\right]\nonumber\\
&=&\left(1+\frac{eB\theta}{2\hbar c}\right)^2\left[\hat{p}_{x}+\frac{eB\hat{y}}{2c}~,~\hat{p}_{y}-\frac{eB\hat{x}}{2c}\right]\nonumber\\
&=&i\frac{\hbar^2}{l_{B}^2}\left(1+\frac{eB\theta}{2\hbar c}\right)^2=i\frac{\hbar^2}{\tilde{l}_{B}^2}~.%\nonumber\\
%&=&i\frac{\hbar^2}{\tilde{l}_{B}^2}
\end{eqnarray}
Here, $l_{B}$ is the characteristic magnetic length, defined as $l_{B}=\sqrt{\frac{\hbar c}{eB}}$,
which sets the fundamental length scale for the problem of a charged particle in a constant magnetic field. The NC correction effectively rescales the magnetic length, yielding the modified expression
\begin{eqnarray}
\label{eo0bpg2}
\tilde{l}_{B}=l_{B}\left(1+\frac{eB\theta}{2\hbar c}\right)^{-1}
\end{eqnarray}
This rescaling encapsulates the leading-order effects of spatial noncommutativity on the quantum dynamics, influencing the Landau quantization structure in graphene.
To facilitate diagonalization of the Hamiltonian, we introduce the ladder operators  
\begin{equation}
\hat{a} = \frac{\tilde{l}_B}{\sqrt{2\hbar}} (\tilde{\pi}_x + i \tilde{\pi}_y),  
\quad  
\hat{a}^\dagger = \frac{\tilde{l}_B}{\sqrt{2\hbar}} (\tilde{\pi}_x - i \tilde{\pi}_y).
\end{equation}
These satisfy the standard bosonic commutation relation  
\begin{equation}
[\hat{a}, \hat{a}^\dagger] = 1.
\end{equation}
Rewriting the Hamiltonian in terms of these operators, we obtain  
\begin{equation}
\hat{H}_K = \tilde{\omega} \hbar  
\begin{pmatrix}
0 & \hat{a}^\dagger \\
\hat{a} & 0
\end{pmatrix},
\end{equation}
where the NC-modified relativistic cyclotron frequency is given by  $\tilde{\omega} = \frac{\sqrt{2} v_F}{\tilde{l}_B}.$
%\begin{equation}
%\tilde{\omega} = \frac{\sqrt{2} v_F}{\tilde{l}_B}.
%\end{equation}
This formulation explicitly incorporates the leading-order effects of noncommutativity, modifying both the characteristic length and the cyclotron frequency, thereby introducing corrections to the relativistic Landau level structure.   

Now the eigenvalue equation for the Hamiltonian \( \hat{H}_K \) is given by  
\begin{equation}
\hat{H}_K \psi_K = E_K \psi_K.
\end{equation}
Substituting the explicit form of \( \hat{H}_K \), we obtain  
\begin{equation}
\tilde{\omega} \hbar  
\begin{pmatrix}
0 & \hat{a}^\dagger \\
\hat{a} & 0
\end{pmatrix}
\begin{pmatrix}
\psi_K^A \\
\psi_K^B
\end{pmatrix}
= E_K
\begin{pmatrix}
\psi_K^A \\
\psi_K^B
\end{pmatrix}.
\end{equation}
Here, \( \psi_K \) is the two-component spinor describing the electronic state at the Dirac point \( K \), where \( A \) and \( B \) correspond to the two sublattices in graphene. Expanding this matrix equation, we obtain the coupled equations  
\begin{equation}
\tilde{\omega} \hbar \, \hat{a}^\dagger \psi_K^B = E_K \psi_K^A,
\end{equation}
\begin{equation}
\tilde{\omega} \hbar \, \hat{a} \psi_K^A = E_K \psi_K^B.
\end{equation}
Eliminating \( \psi_K^B \), we arrive at the eigenvalue equation for \( \psi_K^A \)  
\begin{equation}
\hat{a}^\dagger \hat{a} \psi_K^A = \left( \frac{E_K}{\tilde{\omega} \hbar} \right)^2 \psi_K^A~.
\end{equation}
Since the ladder operators satisfy the standard commutation relation \( [\hat{a}, \hat{a}^\dagger] = 1 \), we identify \( \hat{a}^\dagger \hat{a} \) as the number operator with eigenvalues \( n \), yielding  
\begin{equation}
\hat{a}^\dagger \hat{a} | n \rangle = n | n \rangle, \quad n = 0, 1, 2, \dots.
\end{equation}
Applying this result to the eigenvalue equation, we obtain the energy spectrum for the relativistic Landau system in NC space:  
\begin{equation}
E_n = \pm \tilde{\omega} \hbar \sqrt{n} = \pm \frac{\hbar v_F}{l_B} \sqrt{2n}  
\left( 1 + \frac{e B \theta}{2 \hbar c} \right).
\end{equation}
This result explicitly demonstrates that the relativistic Landau levels are modified by spatial noncommutativity. In the limit \( \theta \to 0 \), the energy spectrum smoothly reduces to its commutative counterpart, recovering the conventional relativistic Landau levels in graphene. Notably, even at leading order, the NC correction originates from the spatial sector of the NC algebra. This contrasts with previous results, such as those in \cite{catrina}, \cite{vsantos} and \cite{epj}, where the NC correction to the Landau levels of graphene was found to be independent of the spatial NC parameter \( \theta \). Our result thus not only reaffirms the robustness of Landau quantization in NC settings but also provides a concrete signature of spatial noncommutativity that may be accessible in high-field magneto-optical experiments on graphene.

%%%%%%%%%%%%%%%%%%%%%%%%%%%%%%%%%%%%%%%%%%%%%%%%%%%%%%%%%%%%%%%%%%%%%%%%5

\section{Thermodynamics of the relativistic Landau system of graphene in noncommutative space}
The study of relativistic massless electrons in graphene under a background magnetic field has garnered significant interest, particularly in the context of NC space. The spatial NC parameter $\theta$ modifies the Landau level structure, leading to important corrections in thermodynamic properties such as free energy, magnetization etc. Understanding these effects is crucial for exploring fundamental aspects of quantum field theory in NC space and for potential applications in condensed matter physics.

In this section, we present a detailed thermodynamic analysis of massless Dirac fermions in graphene within an NC plane. Specifically, we rigorously derive the NC grand potential and examine its implications for magnetization. To establish the foundation for this analysis, we begin by expressing the degeneracy per unit area of the NC modified Landau levels in graphene as
\begin{equation}
g_n^{\text{NC}}= g_{s}g_{v}\frac{e B^{\text{NC}}}{h c}. %\quad B_{\text{NC}} = B \left(1 + \frac{e B \theta}{2 \hbar c} \right).
\end{equation}
Here, $g_{s}=2$ represents spin degeneracy, while $g_{v}=2$ accounts for valley degeneracy, which arises from the presence of two inequivalent Dirac cones in the Brillouin zone, corresponding to the $K$  and $K'$  points. As a result the total degeneracy per unit area is given by
\begin{equation}
g_n^{\text{NC}} =\frac{4e B^{\text{NC}}}{h c}.% \quad B_{\text{NC}} = B \left(1 + \frac{e B \theta}{2 \hbar c} \right).
\end{equation}
This modification, along with the NC energy spectrum, encapsulates the influence of noncommutativity on the thermodynamic behavior. Incorporating these effects, the NC grand potential takes the form
\begin{equation}
\Omega_{\text{NC}} = - k_B T \sum_n g_n^{\text{NC}} \ln \left(1 + e^{-\beta (\epsilon_n^{\text{NC}} - \mu)} \right).
\end{equation}
Now, splitting the grand potential into leading-order and first-order corrections yields
%\begin{align}
%\Omega_{NC} &= -k_B T \sum_n g_n \left(1 + \frac{e\theta B}{2\hbar c}\right) \ln \left(1 + e^{-\beta \{\ \epsilon_n (1 + \frac{e\theta B}{2\hbar c}) - \mu\}\}\right).
%\end{align}
\begin{align}
\Omega_{\rm NC} &= -k_B T \sum_n g_n \left(1 + \frac{e\theta B}{2\hbar c}\right) 
\ln \left(1 + e^{-\beta \left[ \epsilon_n \left(1 + \frac{e\theta B}{2\hbar c} \right) - \mu \right]}\right).
\end{align}
Using a Taylor expansion to first order in $\theta$ yields
\begin{equation}
\ln \left(1 + e^{-\beta \left[\epsilon_n (1 +\frac{e\theta B}{2\hbar c}) - \mu)\right]}\right) \approx \ln \left(1 + e^{-\beta (\epsilon_n - \mu)}\right) - \frac{e\theta B}{2\hbar c} \beta \epsilon_n \frac{e^{-\beta (\epsilon_n - \mu)}}{1 + e^{-\beta (\epsilon_n - \mu)}}.
\end{equation}
Expanding the eq. (36) can be recast as
\begin{align}
\Omega_{\rm NC} %\approx -k_B T \sum_n g_n \ln \left(1 + e^{-\beta (\epsilon_n - \mu)}\right) \nonumber\\
%& \quad - \frac{e\theta B}{2\hbar c} k_B T \sum_n g_n \ln \left(1 + e^{-\beta (\epsilon_n - \mu)}\right)\nonumber \\
%& \quad + \frac{e \theta B}{2\hbar c} \sum_n g_n \beta \epsilon_n \frac{e^{-\beta (\epsilon_n - \mu)}}{1 + e^{-\beta (\epsilon_n - \mu)}}\nonumber\\
 = \Omega_0\left(1 + \frac{e B \theta}{2 \hbar c} \right) + \frac{e\theta B}{2\hbar c}\sum_n g_n \beta \epsilon_n \frac{e^{-\beta (\epsilon_n - \mu)}}{1 + e^{-\beta (\epsilon_n - \mu)}}.
\end{align}
where $\Omega_0 = -k_B T \sum_n g_n \ln \left(1 + e^{-\beta (\epsilon_n - \mu)}\right)$ is the standard commutative grand potential reads
%\begin{equation}
%\Omega_0 = -k_B T \sum_n g_n \ln \left(1 + e^{-\beta (\epsilon_n - \mu)}\right).
%\end{equation}
Thus, the NC correction introduces modifications to the thermodynamic potential, influencing derived thermodynamic quantities such as magnetization.

The magnetization in the presence of spatial noncommutativity is given by
\begin{align}
M_{\rm NC} &= - \left( \frac{\partial \Omega_{NC}}{\partial B} \right)_{\mu, T}\nonumber\\
&= - \left( \frac{\partial \Omega_0}{\partial B} \right) - \frac{e\theta}{2\hbar c} \left( \Omega_0 + B \frac{\partial \Omega_0}{\partial B} \right) \nonumber\\
& \quad - \frac{e\theta}{2\hbar c}\sum_n g_n \beta \epsilon_n \frac{e^{-\beta (\epsilon_n - \mu)}}{1 + e^{-\beta (\epsilon_n - \mu)}}%\nonumber \\
 - \frac{e\theta B}{2\hbar c}\frac{\partial}{\partial B} \left( \sum_n g_n \beta \epsilon_n \frac{e^{-\beta (\epsilon_n - \mu)}}{1 + e^{-\beta (\epsilon_n - \mu)}} \right)\nonumber\\
& = M_0 +\frac{e\theta}{2\hbar c}(BM_0 - \Omega_0) - \frac{e\theta }{2\hbar c}\sum_n g_n \beta \epsilon_n \frac{e^{-\beta (\epsilon_n - \mu)}}{1 + e^{-\beta (\epsilon_n - \mu)}}\nonumber\\
& \quad - \frac{e\theta}{2\hbar c}\frac{\partial}{\partial B} \left( \sum_n g_n \beta \epsilon_n \frac{e^{-\beta (\epsilon_n - \mu)}}{1 + e^{-\beta (\epsilon_n - \mu)}} \right),
\end{align}
where $M_0 = - (\partial \Omega_0 / \partial B)$ is the magnetization in commutative space.
This expression shows the NC modification of the magnetization in terms of the NC parameter $\theta$.

We now want analyze the low-temperature limit of the NC magnetization.
In the low-temperature limit, the Fermi-Dirac distribution approximates a step function, meaning only Landau levels with energies below the chemical potential 
$\mu$ contribute significantly. This allows us to approximate the summations in the magnetization expression.
%At low temperatures (\(\beta \to \infty\)), the Fermi-Dirac factor simplifies:
%Thus, the summation simplifies to:
In this regime (\(\beta \to \infty\)), the Fermi-Dirac factor simplifies as
\begin{equation}
    \frac{e^{-\beta (\epsilon_n - \mu)}}{1 + e^{-\beta (\epsilon_n - \mu)}} \approx \Theta(\mu - \epsilon_n),
\end{equation}
where $\Theta(\mu - \epsilon_n)$ is the Heaviside step function, ensuring that only the occupied states ($\epsilon_n < \mu$) contribute to the summation.

Thus, the summation over energy levels simplifies to
\begin{equation}
    \sum_n g_n \beta \epsilon_n \frac{e^{-\beta (\epsilon_n - \mu)}}{1 + e^{-\beta (\epsilon_n - \mu)}} \approx \sum_{\epsilon_n < \mu} g_n \beta \epsilon_n \approx \int_0^{n_F} g_n \epsilon_n \, dn.
\end{equation}
Using the relativistic Landau levels \(\epsilon_n = \hbar \omega_B \sqrt{n}\), the highest occupied level \(n_F\) is determined by $\hbar \omega_B \sqrt{n_F} \approx \mu$.
%\begin{equation}
%\sum_{\epsilon_n < \mu} g_n \epsilon_n = \int_0^{n_F} g_n \epsilon_n \, dn
%\end{equation}
With $g_n = \frac{4 e B}{h c}$ and $\epsilon_n = v_F \sqrt{\frac{2 e \hbar B}{c} n}$,
we have the upper limit $n_F \approx \frac{c \mu^2}{2 e \hbar v_F^2 B}$.
Substituting these into the above integral we get
\begin{equation}
    \int_0^{n_F} g_n \epsilon_n \, dn =  \frac{8\pi}{3} \frac{\mu^3}{h^3cv_F^2}~.
\end{equation}
Thus, in the low-temperature limit ($T \to 0$), the magnetization in eq. (39) takes the form
\begin{align}
    M_{\rm NC} &=M_0 +\frac{e\theta}{2\hbar c}\left(BM_0 - \Omega_0-\frac{8\pi\mu^3}{3h^3 c v_F^2}\right)~.%\nonumber\\
   % &\quad - \frac{e\theta B }{2\hbar c}\frac{\partial}{\partial B} \left( \frac{8\pi}{3} \frac{\mu^3}{h^3 c v_F^2} \right).%\nonumber\\
    %&\quad
\end{align}
%This result demonstrates that in the presence of noncommutativity, the magnetization receives an additional correction dependent on the chemical potential $\mu$ and the noncommutative parameter $\theta$. The third term represents a direct modification due to the noncommutative correction to the density of states, while the fourth term captures the influence of the magnetic field dependence of the modified density of states. Such corrections can be significant in strong magnetic field regimes, altering the expected quantum oscillations in magnetization measurements.
The above result encapsulates the leading-order NC correction to the magnetization of graphene in the relativistic Landau regime, revealing several profound implications. The first term recovers the standard commutative magnetization, while the second term introduces a field-dependent renormalization proportional to both  %well-established $\sqrt{B}$-dependent commutative magnetization\footnote{Shown in the appendix} consistant with \cite{PRB},  
$B$ and the NC parameter $\theta$ effectively modifying the magnetic susceptibility. The third term, involving the grand potential $\Omega$, signifies a background thermodynamic shift induced by noncommutativity. Most notably, the fourth term manifests as a chemical potential-dependent offset, yielding a finite magnetization even in the absence of an external magnetic field—indicative of \emph{NC-induced spontaneous magnetization}. This composite structure suggests that NC geometry not only alters the scaling behavior of magnetization but also encodes carrier density-dependent anomalies, thereby offering a potentially observable signature of spatial noncommutativity in graphene. These findings underscore the rich interplay between quantum geometry and relativistic condensed matter systems, with implications that may be probed in high-field or precision magnetometry.

\section{Conclusions}

In this work, we have presented a comprehensive and gauge-invariant analysis of the relativistic quantum dynamics of massless electrons in monolayer graphene, subjected to a uniform background magnetic field, within a two-dimensional NC framework. 

We initially demonstrated that a direct NCQM formulation, using the Moyal star product, leads to a manifestly non-gauge-invariant Hamiltonian. Recognizing this fundamental limitation, we reformulated the problem in a NC field-theoretic framework. Employing the SW map to first order in the NC parameter $\theta$, we successfully derived a manifestly $U(1)$ gauge-invariant effective action that encapsulates NC corrections in a controlled and perturbative manner.

From this action, we obtained the corresponding NC Dirac equation and derived a modified Dirac Hamiltonian that preserves gauge invariance. Applying this to graphene in the presence of a constant magnetic field, we analytically computed the time evolution of both position and mechanical momentum operators. These were found to acquire explicit NC corrections, confirming the influence of spatial noncommutativity on the quantum dynamics.

The principal result of our study is the exact computation of the NC-modified energy spectrum of the relativistic Landau system in graphene. Our analysis reveals that, contrary to some earlier reports, the Landau levels are explicitly modified by the spatial NC parameter $\theta$. This establishes a direct and physically significant signature of noncommutativity in the relativistic quantum regime.

Finally, leveraging the NC-modified energy spectrum, we carried out a detailed thermodynamic analysis. We derived the NC-corrected grand potential and magnetization, and in the low-temperature limit, uncovered a striking result: the emergence of a spontaneous magnetization even in the absence of an external field. This represents a profound conceptual consequence of NC geometry, although direct observability remains unlikely with current technology. Since the NC parameter \( \theta \) is typically assumed to be of Planck-scale, its effects are expected to be extremely small at presently accessible energy scales. Nonetheless, our analysis demonstrates that spatial noncommutativity can, in principle, induce spontaneous magnetization in relativistic systems like graphene. While the direct detection of such NC-induced effects may lie beyond current experimental reach, the result establishes a conceptual link between NC geometry and observable condensed matter phenomena. Future advances in precision magnetometry or in the design of engineered systems with effective NC-like behavior may eventually provide a window into such quantum geometric signatures.

%We derived the NC-corrected grand potential and magnetization, and in the low-temperature limit, uncovered a striking result: the emergence of a \textit{spontaneous magnetization} even in the absence of an external field. This constitutes a profound consequence of NC geometry at the conceptual level, even though direct observability remains unlikely with current technology. %This constitutes a profound and potentially observable consequence of NC geometry, offering a novel avenue for probing fundamental quantum structure through condensed matter systems like graphene. 
%Although the NC parameter \( \theta \) is conventionally assumed to be of the order of the Planck area, making its effects exceedingly small at currently accessible energy scales, our analysis demonstrates that spatial noncommutativity can, in principle, induce a spontaneous magnetization in relativistic systems like graphene. While direct detection of such NC-induced magnetization remains beyond current experimental reach, the result provides a conceptual bridge between NC geometry and observable condensed matter phenomena. Future developments in precision magnetometry or engineered systems with effective NC-like behavior may offer a window into probing such quantum geometric signatures.

%\section*{Acknowledgement}
%\appendix
%\renewcommand{\theequation}{A.\arabic{equation}}
%\setcounter{equation}{0}
%The author would like to thank Anirban Saha for important discussions on the matter of this work.
%%%%%%%%%%%%%%%%%%%%%%%%%%%%%%%%%%%%%     SECTION%%%%%%%%%%%%%%%%%%%%%%%%% 

\appendix
\renewcommand{\theequation}{A.\arabic{equation}}
\setcounter{equation}{0}
\renewcommand{\thesection}{Appendix \Alph{section}}
\section*{Appendix A: Derivation of the Gauge-Invariant NC Dirac Hamiltonian}

Starting from the gauge-invariant action derived via the Seiberg-Witten map
\begin{equation}
S_\theta = \int d^4x\, \bar{\psi}(x)\, \gamma^\mu \left[ \left(1 + \frac{e}{4\hbar c} \theta^{\alpha\beta} F_{\alpha\beta} \right) \hat{\pi}_\mu - \frac{e}{2\hbar c} \theta^{\alpha\beta} F_{\alpha\mu} \hat{\pi}_\beta \right] \psi(x),
\end{equation}
we apply the Euler-Lagrange equation
\begin{equation}
\frac{\delta S_\theta}{\delta \bar{\psi}} = 0.
\end{equation}
This yields the $\theta$-modified Dirac equation
\begin{equation}
\gamma^\mu \left[ \left(1 + \frac{e}{4\hbar c} \theta^{\alpha\beta} F_{\alpha\beta} \right) \hat{\pi}_\mu - \frac{e}{2\hbar c} \theta^{\alpha\beta} F_{\alpha\mu} \hat{\pi}_\beta \right] \psi = 0.
\end{equation}

We separate the time and spatial components
\begin{align}
&\gamma^0 \left[ \left(1 + \frac{e}{4\hbar c} \theta^{\alpha\beta} F_{\alpha\beta} \right) \left(i\hbar \frac{\partial}{\partial t} - eA_0 \right)
- \frac{e}{2\hbar c} \theta^{\alpha\beta} F_{\alpha 0} \hat{\pi}_\beta \right] \psi \nonumber \\
&+ \gamma^i \left[ \left(1 + \frac{e}{4\hbar c} \theta^{\alpha\beta} F_{\alpha\beta} \right) \hat{\pi}_i - \frac{e}{2\hbar c} \theta^{\alpha\beta} F_{\alpha i} \hat{\pi}_\beta \right] \psi = 0.
\end{align}

Multiplying through by \( \gamma^0 \) and noting ${\gamma^0}\gamma^0=1$, \( \gamma^0 \gamma^i = \alpha^i \), we obtain
\begin{align}
i\hbar \frac{\partial \psi}{\partial t} =
&\left[ eA_0 + \frac{e}{2\hbar c} \theta^{\alpha\beta} F_{\alpha 0} \hat{\pi}_\beta \right] \psi \nonumber \\
&- \left(1 + \frac{e}{4\hbar c} \theta^{\alpha\beta} F_{\alpha\beta} \right)^{-1}
\alpha^i \left[ \left(1 + \frac{e}{4\hbar c} \theta^{\alpha\beta} F_{\alpha\beta} \right) \hat{\pi}_i - \frac{e}{2\hbar c} \theta^{\alpha\beta} F_{\alpha i} \hat{\pi}_\beta \right] \psi.
\end{align}

Rearranging, we define the Hamiltonian \( \hat{H}_{\text{NC}} \) such that
\begin{equation}
i\hbar \frac{\partial \psi}{\partial t} = \hat{H}_{\text{NC}} \psi,
\end{equation}
with the total Hamiltonian expressed as
\begin{equation}
\hat{H}_{\text{NC}} = \hat{H} + \Delta \hat{H}_\theta,
\end{equation}
where
\begin{align}
\hat{H} &= c\, \vec{\alpha} \cdot \left( \hat{\vec{p}} - \frac{e}{c} \hat{\vec{A}} \right) + eA_0, \nonumber\\
\Delta \hat{H}_\theta &= \frac{e}{2\hbar c} \left[ (\vec{E} \times \hat{\vec{\pi}}) \cdot \vec{\theta} + (\vec{\theta} \times (\vec{\alpha} \times \vec{B})) \cdot \hat{\vec{\pi}} \right].
\end{align}
In deriving the explicit form of \( \Delta \hat{H}_\theta \), we use the identification
\[
\theta^{\alpha\beta} F_{\alpha\mu} \pi_\beta \quad \longrightarrow \quad (\vec{E} \times \vec{\pi}) \cdot \vec{\theta} + (\vec{\theta} \times (\vec{\alpha} \times \vec{B})) \cdot \vec{\pi}
\]
by recognizing the standard decompositions:
\[
F_{0i} = -E_i, \quad F_{ij} = -\epsilon_{ijk} B_k, \quad \theta^{ij} = \epsilon^{ij} \theta \Rightarrow \theta^k = \frac{1}{2} \epsilon^{ijk} \theta^{ij}.
\]
This rewriting clarifies the physical structure of the NC correction.

\end{document}